\documentclass[12pt]{article}

\begin{document}

\title{Conformal Galilean-type algebras, massless particles and gravitation\thanks{Based on a Plenary Lecture delivered at the workshop ``Sym\`{e}tries non-relativistes: th\`{e}orie math\`{e}matique et applications physiques'', LMPT Tours, 23--24 Juin 2009}}
\author{Peter C. Stichel\\ An der Krebskuhle 21\\
33619 Bielefeld\\
e-mail: peter@physik.uni-bielefeld.de}
\date{July 27, 2009}

\maketitle

\begin{abstract}
After defining conformal Galilean-type algebras for arbitrary dynamical exponent $z$ we consider the particular cases of the conformal Galilei algebra (CGA) and the Schr\"odinger Lie algebra (sch). Galilei massless particles moving with arbitrary, finite velocity are introduced 
\begin{description}
\item{i)} in $d=2$ as a realization of the centrally extended CGA in 6 dimensional phase space,
\item{ii)} in arbitrary spatial dimension $d$ as a realization of the unextended \it{sch} in $4d$ dimensional phase space.
\end{description} 
A particle system, minimally coupled to gravity, shows, besides Galilei symmetry, also invariance with respect to arbitrary time dependent translations and to dilations with $z = (d +2)/3$. The most important physical property of such a self-gravitating system is the appearance of a dynamically generated gravitational mass density of either sign. Therefore, this property may serve as a model for the dark sector of the universe. The cosmological solutions of the corresponding hydrodynamical equations show a deceleration phase for the early universe and an acceleration phase for the late universe. This paper is based, in large part, on a recent work with W.J.~Zakrzewski: Can cosmic acceleration be caused by exotic massless particles? arXiv: 0904.1375 (astro-ph.CO) [1]. 
\end{abstract}

\newpage

\noindent
{\bf Contents}

\begin{description}
\item{1.} Conformal Galilean-type algebras
\item{2.} Galilean massless particles
\item{3.} Coupling to gravity
\item{4.} A self-gravitating fluid
\item{5.} Cosmological solutions
\item{6.} Final remarks
\end{description}

\section{Conformal Galilean-type algebras}

We consider ($d+1$)-dimensional space-time.

\subsection{Conformal subalgebra}

For arbitrary {\bf dynamical exponent $z$} we have in terms of differential operators

$$   H\,=\,\partial_t,\qquad \hbox{\bf Time} \quad \hbox{\bf translations}$$
$$ D\,=\,t\partial_t\,+\,\frac{1}{z}\,x_i\partial_i \quad \hbox{\bf Dilations}$$
$$ K\,=\,t^2\partial_t\,+\,\frac{2}{z}t\,x_i\partial_i\quad \hbox{\bf Expansions} \quad i=1,2...,d.$$
       
Thus $x_i$ and $t$ transform differently with respect to:

\noindent
Dilations
$$x_i^{\star}\,=\,\lambda^{\frac{1}{z}}\,x_i,\qquad t^{\star}\,=\,\lambda t,\qquad \hbox{where}\,\lambda>0,$$

\noindent
Expansions
$$ x_i^{\star}\,=\,\left(\frac{1}{1-kt}\right)^{\frac{2}{z}}x_i,\qquad t^{\star}\,=\,\frac{t}{1-kt},\quad \hbox{where}\,k\in R^1.$$
The operators $(H,D,K)$ form the $O(2,1)$ algebra (independent of $z$)
$$[D,H]\,=\,-H,\quad [K,H]\,=\,-2D,\quad [D,K]\,=\,K.$$

\subsection{ (Enlarged) Galilean algebra}

Define (disregard rotations):
$$ P_i\,=\,-\partial_i,\qquad \hbox{\bf Space}\quad \hbox{\bf translations}$$
$$ K_i\,=\,-t\partial_i \qquad \quad \hbox{\bf Boosts}$$
and the components of the enlargement
$$ F_i\,=\,-t^2\partial_i\qquad \quad \hbox{\bf Accelerations}$$
etc., together, forming the Galilei algebra (considering only nonvanishing Lie brackets)
$$ [H,K_i]\,=\,P_i$$
and its enlargements
$$ [H,F_i]\,=\,2K_i\qquad \hbox{etc.}$$

\subsection{Mixed Lie brackets}

Lie brackets between conformal and (enlarged) Galilei generators depend on $z.$ In particular:
$$[D,P_i]\,=\,-\frac{1}{z}P_i,\quad [D,K_i]\,=\,(1-\frac{1}{z})K_i,$$
$$[K,P_i]\,=\,-\frac{2}{z}K_i,\quad [K,K_i]\,=\,(1-\frac{2}{z})F_i,$$
$$[K,F_i]\,=\,..... \hbox{etc},$$
This infinite series terminates if $z=\frac{2}{n}$ for $n\in N$.

\medskip
\noindent
Particular cases are:
\begin{description}
\item{1.} $z=2$ Schr\"odinger algebra (sch)

We have no accelerations $F_i$. For arbitrary $d$ one central extension is possible
$$[P_i,K_j]\,=\,m\delta_{ij},\qquad m=\hbox{ mass}.$$

\item{2.} $z=1$ Conformal Galilei algebra (CGA) (enlarged by accelerations $F_i$)

We obtain $m=0$ from the Jacobi identity
$$0\,=\,[H,[P_i,F_k]]\,+\,[P_i,[F_k,H]]\,+\,[F_k,[H,P_i]]$$
$$ \quad\quad\,\,\,\,\quad \Downarrow\qquad \quad \qquad \Downarrow \qquad \,\,\qquad \quad \Downarrow$$
\qquad\qquad \qquad\qquad\qquad \, \,\qquad \qquad $c$-number \quad \qquad -2$K_k$\qquad \,\, \quad \quad 0.

\smallskip
\noindent
This argument (for $m=0$) holds for $z\ne2$, {\it ie} if accelerations are present.
\end{description}

\bigskip
\noindent
{\bf Remark:} CGA is the algebra obtained by means of the nonrelativistic contraction from the conformal Poincar\`{e} algebra [4].

\bigskip
\noindent
For $d=2$ we can introduce one central extension:
$$[K_i,K_j]\,=\,\theta\,\epsilon_{ij}.$$

Introducing two additional central charges for the acceleration enlarged Galilean algebra leads to a onefold centrally extended, two
parameter deformed CGA [5].

\bigskip
\noindent
{\bf Remark}: In the expansion-less case we have no accelerations. The algebra consisting of the Galilean generators and dilations closes for arbitrary $z$.

\bigskip
\noindent
{\bf References} for this section:
\begin{description}
\item{1.} Algebra: Henkel [2]; Negro et al, [3]

\item{2.} Central extensions and dynamical realisations in $d=2$; Lukierski, Stichel and Zakrzewski [4,5].
\end{description}

\section{Galilean massless particles}
     
\subsection{Statement}

The realisation of $[P_i,K_j]=0$ ($\Leftrightarrow m=0$) by means of $K_j=f_j$ (phase space 
variables) with $\frac{d}{dt}f_j=0$ requires an enlarged phase space.

To show this consider a massless particle in $d$-dim space with position $x_i$ and canonical momentum $p_i$.

Then proceed in {\bf three steps}:
\begin{description}
\item{1.} {\bf Conserved translations}: We get $[x_i,p_j]=\delta_{ij}$ and $\dot p_j=0$.\\
{\bf Conserved boosts}: From $[x_i,K_j]=t\delta_{ij}$ we obtain $K_j=p_jt+q_j$

with   $\dot q_j=-p_j$ and  $[q_j,x_i]=0$.\\
{\bf The Requirement}

$[p_i,K_j]=0$ leads to $[q_j,p_i]=0$

{\it ie} we obtain
$$ q_j\,\ne\,g_j(x_i,p_i)$$
Thus $\tilde d:=$ dim (phase space) $\ge 3d$.

(cp. to $\tilde d=2d$ in standard case).
\vskip 0.3cm

{\bf Question}: What is $\tilde d_{min}$?

\item{2.} Next we introduce {\bf velocities} $y_i:=[x_i,H]$.

Then, by applying the Jacobi identity, we obtain for 
translations: $[y_i,p_j]=0$

resp. for boosts: $[y_i,K_j]=\delta_{ij}$ leading to $[y_i,q_j]=\delta_{ij}$

and $y_i\ne h_i(x_k,p_k)$ (otherwise we would get $[h_i,q_j]=0$ due to 

 $[q_j,p_i]=[q_j,x_i]=0$).

\item{3.} Now we have to distinguish between the cases $d=2$ and $d\neq 2$. 

\item{a)} In {\bf $d=2$} we can realise $[y_i,q_j]=\delta_{ij}$ by
$$ q_i\,=\,\theta\,\epsilon_{ij}\,y_j\quad \hbox{with}\quad [y_i,y_j]\,=\,\frac{\epsilon_{ij}}{\theta}$$
(leading to $[K_i,K_j]\,=\,\theta \epsilon_{ij}$).

Thus we see that $\tilde d_{min}=6.$

Now let us derive the corresponding minimal 1st-order Lagrangian [4].

$$\mbox{From}~~~ \dot q_i\,=\,-p_i ~~~~\mbox{we obtain}~~~~ \dot y_i\,=\,\frac{1}{\theta}\epsilon_{ij}p_j$$
{\it ie}  EOM are derived by means of PBs from $H=p_iy_i$
being equivalent to
$$L\,=\,p_i\dot x_i\,-\,\frac{\theta}{2}\epsilon_{ij}y_i\dot y_j\,-\,H.$$

\item{b)} In $d\ne 2$ (resp. $\theta = 0$ for $d=2$) we have to assume that $[y_i, q_j] = \delta_{ij}$ together with $[q_i, q_j] = 0$. The latter requirement follows from $[K_i, K_j] = 0$. Clearly the ansatz $q_i = q_i(\vec{y})$ contradicts our assumption. 
Thus $q_i$ must be independent of $y_i$ and so the phase space = $\{x_i,p_i,y_i,q_i\}$ where ${i=1..d}$ and so $\tilde d_{min}=4d$.

Now consider the corresponding minimal 1st-order Lagrangian.

The EOM are
$$ \dot x_i=y_i,\quad \dot p_i=0,\quad \dot q_i=-p_i,\quad \dot y_i=?$$

\noindent
The Poisson brackets are
$$[x_i,p_j]=\delta_{ij},\quad [y_i,q_j]=\delta_{ij}$$
and they imply that
$$ H\,=\,p_iy_i\,+\,f(q_i).$$

 \noindent
The minimal (parameter free) Lagrangian, obtained from $H_{f=0}$ 
is then
$$ L_0\,=\,p_i\dot x_i\,+\,q_i\dot y_i\,-\,p_iy_i.$$
\end{description}

\subsection{Conformal generators}

\begin{description}
\item{a)} {$\bf d=2$} [4].

{\bf Dilation} (conserved)~~~  $D=tH-x_ip_i$

{\bf Expansion} (conserved) $K=-t^2H+2tD-2\theta \epsilon_{ij}x_iy_j$
$$\mbox{We obtain}~~~~~~~~~~~~~~~ [D,P_i]=-P_i$$
{\it ie} we have ${\bf z=1}$  (CGA)

\item{b)} {$\bf d\ne 2$} (resp. $\theta = 0$ for $d=2$)

{\bf Dilation} (conserved for any $z$)  $D=tH-\frac{1}{z}x_ip_i+(1-\frac{1}{z})y_iq_i$

{\bf Expansion} We make the ansatz
 $K=-t^2H+2tD +\alpha x_iq_i.$

Then from $\frac{d}{dt}K=0$ we get $\alpha=-1$ and so {\bf $z=2$}
\end{description}

\bigskip
\noindent
{\bf Remark}: Our massless particles move with arbitrary finite velocity. Therefore they are distinct from the Galilean massless particles introduced by Duval and Horvathy [6] moving with infinite velocity.

\section{Coupling to gravity}

We start, for any $d$, with the parameter-free Lagrangian
$$ L_0\,=\,p_i\dot x_i\,+\,q_i\dot y_i\,-\,H$$
leading to the EOM
\begin{equation}
\ddot x_i\,=\,\dot y_i\,=\,0.
\end{equation}

We introduce a minimal coupling to the gravitational field $\phi(\vec x,t)$ 
$\vec x=\{x_i\},\,(i=1,..d)$ in accordance with Einstein's equivalence principle (``free falling elevator'').

As, at each fixed point $\vec x$, the gravitational force $-\partial_i \phi$ is equivalent to an acceleration
$b_i(t)$ we see that the equation (1) has to be modified to
\begin{equation}
\ddot x_i(t)\,=\,-\partial_i\,\phi(\vec x(t),t)
\end{equation}
because (2) is invariant with respect to arbitrary time-dependent translations (cp. [7])
$$ x_i\,\rightarrow\,x_i'\,=\,x_i\,+\,a_i(t)$$
provided that $\phi$ transforms as
$$\phi(\vec x,t)\,\rightarrow\,\phi'(\vec x',t)\,=\,\phi(\vec x,t)\,-\,\ddot a_ix_i\,+\,h(t).$$

Note that (2) can be realised if we add to $L_0$ an interaction part
$$ L_{int}\,=\,q_i\,\partial_i\phi.$$
Then the EOM $\dot p_i=0$  gets replaced by
$$\dot p_i\,=\,q_k\,\partial_k\partial_i\,\phi.$$

\section{A self-gravitating fluid}

\subsection{Lagrange picture}

We generalise the one-particle phase space coordinates $A_i$ ($A_i\in(x_i,p_i,y_i,q_i)$) to the continuum labeled by
$\vec \xi\in R^d$ (comoving coordinates)
$$A_i(t)\,\rightarrow \,A_i(\vec \xi,t).$$

The Lagrangian $L$ for the self-gravitating fluid becomes
$$L=\int d^d\xi \left[L_0(A_i(\vec \xi,t))+L_{int}(A_i(\vec \xi,t),\phi(\vec x(\vec \xi,t)))\right] \,+\,L_{\phi}$$
where the field part $L_{\phi}$ has the standard form
$$L_{\phi}\,=\,-\frac{1}{8\pi G}\,\int\,d^dx\,(\partial_i\,\phi)^2,$$
and $G$ is Newton's gravitational constant.

\subsection{Eulerian picture}

We transform $A_i\in (p_i,y_i,q_i)$ from comoving coordinates $\vec\xi$ to fixed space coordinates $\vec x\in R^d$,
{\it ie} $A_i(\vec \xi,t)\rightarrow A_i(\vec x,t)$ by (cp. [8])
$$A_i(\vec x,t)n(\vec x,t)\,=\,\int d^d\xi\,A_i(\vec \xi,t)\,\delta(\vec x-\vec x(\vec \xi,t)).$$
\vskip -0.2cm
Here $n(\vec x,t)$ is the particle number density 
$$ n(\vec x,t)\,:=\,\int d^d\xi\,\delta(\vec x-\vec x(\vec \xi,t)).$$

We apply these transformations to the Euler-Lagrange EOM obtained from $L$ in section 4.1 
(def. $u_i(\vec x,t):=y(\vec x,t)$) and get the
{\bf fluid dynamical EOM}
$$ \partial_t n\,+\,\partial_k(n\,u_k)\,=\,0,\quad \hbox{\bf Continuity\,\,equation}$$
$$ D_t\,u_i\,=:\,-\partial_i\phi\quad \quad \hbox{\bf Euler\,\,equation}\qquad \hbox{where}$$
\vskip -0.6cm
$$ D_t\,:=\,\partial_t\,+\,u_k\partial_k\qquad \hbox{(convective\,\,derivative)}$$
and
$$\triangle \phi\,=\,4\pi\,G\,\partial_k (nq_k)\quad \quad \hbox{\bf Poisson\,\,equation}.$$
Note that the right hand side of the Poisson equation describes a dynamically 
generated active gravitational mass density (in standard terms given by ``energy density + 3 pressure'') 
which may be of either sign

\smallskip
\noindent
+ sign leading to  {\bf attractive} gravitation

\smallskip
\noindent
- sign leading to {\bf repulsive} gravitation.

\smallskip
\noindent
This promotes the self-gravitating fluid to a possible candidate for the dark sector of the universe!

To get the $q_i$ we have to solve the additional EOM - due to the enlarged phase space -{\it ie}
$$D_t\,q_i\,=\,-p_i,\quad D_t\,p_i\,=\,q_k\,\partial_k\partial_i\phi.$$

\subsection{Symmetries}

Fluid dynamical EOM exhibit the following symmetries:

\begin{description}
\item{i)} Clearly {\bf rotational symmetry}

\item{ii)} Invariance with respect to infinitesimal {\bf time-dependent translations}
$$ \delta x_i\,=\,a_i(t)\qquad \hbox{with}$$
$$ \delta \phi\,=\,-\ddot a_i x_i\,-\,a_k\partial_k \phi\,+\,h(t)$$
$$ \delta u_i\,=\,\dot a_i\,-\,a_k\partial_k\phi.$$
All other fields transform as scalars, {\it ie}
$$ \delta n\,=\,-\,a_k\partial_k n\quad \quad (etc)$$
(the same for $p_i$, resp. $q_i$).
\end{description}

Note that by a suitable choice of a time dependent translation one can pass to a Galilean frame in which the solution of the Poisson eq. is given by (in the {\bf $d=3$} case)
$$\phi(\vec x,t)\,=\,-G\,\int\,d^3x'\,\frac{(\partial_i(nq_i))(\vec x',t)}{\vert \vec x-\vec x'\vert}.$$

Then the {\bf symmetry algebra} becomes the expansion-less conformal Galilei algebra with $z=\frac{5}{3}$
($z=\frac{d+2}{3}$ for any $d$).

{\bf Conserved Generators} are:
$$P_i\,=\,\int \,d^3x\, n\,p_i\quad \quad \hbox{\bf Linear \,\, momentum}$$
$$K_i\,=\,tP_i\,+\,Q_i\qquad \qquad \hbox{\bf Boost \,\,generators}$$
with 
$$ Q_i\,:=\,\int \,d^3x\,n\,q_i$$\\
Note that we still have
$$[P_i,K_j]\,=\,0.$$

We also have
$$\vec J\,=\,\int\,d^3x\,n\,(\vec x\wedge \vec p\,+\,\vec u\wedge\vec q)\quad \hbox{\bf Angular \,\,momentum}$$
and {\bf Energy} is given by
$$H=\int d^3x n\,p_i u_i\,-\,\frac{G}{2}\int d^3x\,d^3x'\,\frac{(\partial_i(nq_i))(\vec x,t)(\partial_j(nq_j))(\vec x',t)}
{\vert \vec x-\vec x'\vert}.$$

We make the following ansatz for the {\bf dilation charge}
$$D=Ht\,+\,(1-\frac{1}{z})\int d^3x\,n q_i u_i\,-\,\frac{1}{z}\int d^3x\, n \,x_i\,p_i$$
which, from $\frac{d}{dt}D=0$, gives $z=\frac{5}{3}.$

\bigskip
\noindent
{\bf Remark:} Adding standard matter (eg. cold dark matter cp. [1]) leads to the violation of dilation symmetry because the corresponding kinetic term would scale with $z=2$ whereas the potential $(1/r)$ term would scale with $z=1$. 

\section{Cosmological solutions}

We consider the self-gravitating fluid as a model for the dark sector of the universe. Hence 
we look for solutions of the fluid EOM satisfying the cosmological principle (the universe  
is supposed to be isotropic and homogeneous on large scales).

This gives us (in a suitable Galilean frame)
$$n(\vec x,t)\,=\,n(t),\quad u_i(\vec x,t)\,=\,x_i\,H(t).$$

Here $H(t)$ is the {\bf Hubble ``parameter''} given in terms of the {\bf cosmic scale factor} $a(t)$ by
$$H\,=\,\frac{\dot a}{a}$$
and
$$ q_i(\vec x,t)\,=\,x_i\,g(a(t)).$$

We insert all this into the EOM and get:

\begin{description}
\item{i)} From the continuity eq.
$$\dot n\,+\,3\frac{\dot a}{a}n\,=\,0 ~~~\hbox{leading, after integration, to}~~~ n(t)\,=\,\frac{D}{\frac{4\pi}{3}a^3(t)}$$
with $D$=const $>0$.

\item{ii)} From the Euler eq.
$$ \phi(\vec x,t)\,=\,\frac{r^2}{2}\varphi(t) ~~~\mbox{with}~~~ \varphi(t)\,=\,-\frac{\ddot a}{a}.$$

\item{iii)} From the EOM for $p_i$ and $q_i$
$$ \ddot g\,+\,2\frac{\dot a}{a}\dot g\,=\,0$$
which, after integration, gives us
\begin{equation} \dot g(a(t))\,=\,\frac{\beta}{a^2(t)},\quad \beta=\hbox{const}.
\end{equation}

\item{iv)} Putting all this into the Poisson eq. we get
\begin{equation}
 -\ddot a\,=\,\frac{3GD}{a^2}\,g(a)
\end{equation}
{\it ie} a {\bf Friedmann-like equation}.
\end{description}

Using (3) we integrate (4) twice obtaining a cubic equation for $g(a)$
\begin{equation}
g(g^2+C_1)\,+\,C_0\left(1-\frac{a_t}{a}\right)\,=\,0,
\end{equation}
where $C_{0,1}$ are integration constants and
$$a_t\,:=\,\frac{2\beta^2}{GDC_0}.$$

If now we have $C_{0,1}>0$ and we use $a(t)$ as a measure of time (we have $\dot a>0$, {\it ie}
an expanding universe) we get from (5) that
\begin{itemize}
\item For $a<a_t$ we have $g(a)>0$ and so from (4) $\ddot a<0$ 
{\it ie} we are in the deceleration phase of the early universe.
\item For $a>a_t$ we get from (5) $g(a)<0$, hence $\ddot a>0$ and we are in the 
acceleration phase of the late universe.
\end{itemize}

Hence $a_t$ defines the point of transition from one phase to the other. This picture is consistent
with astrophysical observations.

\bigskip
\noindent
{\bf Remark:} $C_{0,1}>0$ is also necessary to obtain these results. 

Furthermore, $C_{0,1}=0$ would reproduce the scale invariant
solution valid asymptotically at small $t$ (then $a(t)\sim t^{\frac{3}{5}}$).

To prove this we start with the behaviour of the particle density $n(\vec{x},t)$ with respect to dilations
$$
[n(\vec{x},t),D] = (t\partial_t + \frac{3}{5} x_i \partial_i + \frac{9}{5}) n(\vec{x},t)
$$
{\it ie} that the scale invariant solution for $n(t)$ is given by
$$
n(t) \sim t^{-9/5}
$$
and from
$$ n(t) \sim a^{-3} (t)
$$
we obtain the stated result.

\section{Final remarks}

\subsection{The main achievements of our model are}

\begin{itemize}

 \item Parameter free Lagrangian for a self-gravitating system of Galilean massless particles
 \item Dynamical generation of an active gravitational mass density of either sign
\item Explanation of the deceleration phase of the early universe and of the acceleration phase of the late universe
 
 \end{itemize}

In order to see clearly the relevance of our results we have to contrast our model with other attempts to explain the accelerated expansion of the universe (for details and references see
 [1]).

Other models, introduced in the framework of General Relativity, consist of two classes:

\begin{description}
\item{i)} either one modifies the geometric part of the Einstein-Hilbert action,
\end{description}
or
\begin{description}
\item{ii)} one modifies the matter part of this action by introducing either a positive cosmological constant or a dynamical model leading to negative pressure (scalar fields etc.).
\end{description}

But all these models contain at least one new parameter (in most cases even some unknown function). None of these models is derived from fundamental physics. This distinguishes other models from ours.

\subsection{Open problems (cosmological solutions)}

\begin{itemize}

 \item Determination (or restriction), by 
physical arguments,  of the a-priori unknown integration constants, arising from the additional phase space dimensions
\item Comparison with astrophysical data

But such a comparison meets the difficulty that the determination of cosmological parameters from observational data is at present mostly model dependent (cp. [9] for the case of $a_t$). 
 \end{itemize}

\subsection{Open problems of a general nature} 

Relativistic generalisation of our model and its relation to the framework of General Relativity.

It is unlikely, if not impossible, to obtain our model as a nonrelativistic limit of a relativistic model. Massless relativistic particle models  show conformal Poincar\`{e} symmetry leading, in the nonrelativistic limit, to conformal Galilei symmetry [4], {\it ie} $z=1$. But we have $z=5/3$. In [1] we have speculated, that the relativistic generalization of our Galilean massless particles are tachyons. But it seems to me more likely that we meet here a situation being to Ho\v{r}ava gravity [10], where we have nonrelativistic symmetry in the ultraviolet (small $t$) and approach  General Relativity only in the infrared limit (large $t$). To establish such a picture we have to modify our model in an appropriate manner. This is a challenge for further research. 

\bigskip
\medskip
\noindent
{\bf Acknowledgement:} I'm grateful to Wojtek Zakrzewski for a critical reading of the manuscript and some helpful comments. 

\section*{References}

\begin{description}
\item{[1]} P.C.~Stichel and W.J.~Zakrzewski, Can cosmic acceleration be caused by exotic massless particles?, arXiv: 0904.1375 (astro-ph.CO).

\item{[2]} M.~Henkel, Phys.~Rev.~Lett.~{\bf 78}, 1940 (1997).

\item{[3]} J.~Negro et al, J.~Math.~Phys.~{\bf 38}, 3786 (1997)

\item{[4]} J.~Lukierski, P.C.~Stichel and W.J.~Zakrzewski, Phys.~Lett.~{\bf A 357}, 1 (2006).

\item{[5]} J.~Lukierski, P.C.~Stichel and W.J.~Zakrzewski, Phys.~Lett.~{\bf B 650}, 203 (2007).

\item{[6]} C.~Duval and P.~Horvathy, arXiv: 0904.0531 (math-ph).

\item{[7]} O.~Heckmann and E.~Sch\"ucking, Z.~Astrophys.~{\bf 38}, 95 (1955) (in German).

\item{[8]} R.~Jackiw et al, J.~Phys.~{\bf A 37}, R 327 (2004).

\item{[9]} Lixin Xu et al, arXiv: 0905.4552 (astro-ph.CO).

\item{[10]} P.~Ho\v{r}ava, Phys.~Rev.~{\bf D 79}, 084008 (2009); arXiv: 0901.3775 (hep-th).
\end{description}

\end{document}